%% file: WaldOptimality_ISIT17_corrected.tex
\pdfoutput=1

\documentclass[a4paper,conference]{IEEEtran}
\IEEEoverridecommandlockouts

\usepackage[utf8]{inputenc}
\usepackage[T1]{fontenc}
\usepackage{url}
\usepackage{ifthen}
\usepackage[cmex10]{amsmath} 
\usepackage{cite}
\usepackage{psfrag,graphicx}
\usepackage[tight,footnotesize]{subfigure}
\usepackage[cmex10]{amsmath}
\usepackage{amsthm}
\usepackage{bbm}
\usepackage{vmr-symbols-vecbold}
\usepackage{standard-macros}
\usepackage{subfigure}

\definecolor{Red}{rgb}{1.0, 0, 0}

\newtheorem{theorem}{Theorem}
\newtheorem{corollary}{Corollary}
\newtheorem{lemma}{Lemma}

\begin{document}

\title{An Information Theoretic Analysis\\of Sequential Decision-Making}

\author{
\IEEEauthorblockN{Meik D\"orpinghaus\IEEEauthorrefmark{1}\IEEEauthorrefmark{4}, \'{E}dgar Rold\'{a}n\IEEEauthorrefmark{2}\IEEEauthorrefmark{4}\IEEEauthorrefmark{6}, Izaak Neri\IEEEauthorrefmark{2}\IEEEauthorrefmark{3}\IEEEauthorrefmark{4}, Heinrich~Meyr\IEEEauthorrefmark{1}\IEEEauthorrefmark{4}\IEEEauthorrefmark{5}, and Frank J\"ulicher\IEEEauthorrefmark{2}\IEEEauthorrefmark{4}}
\IEEEauthorblockA{\IEEEauthorrefmark{1}Vodafone Chair Mobile Communications Systems, Technische Universit\"at Dresden,\\ Dresden, Germany, meik.doerpinghaus@tu-dresden.de}
\IEEEauthorblockA{\IEEEauthorrefmark{2}Max-Planck-Institute for the Physics of Complex Systems, Dresden, Germany, \{edgar, izaak, julicher\}@pks.mpg.de}
\IEEEauthorblockA{\IEEEauthorrefmark{3}Max-Planck-Institute of Molecular Cell Biology and Genetics, Dresden, Germany}
\IEEEauthorblockA{\IEEEauthorrefmark{4}Center for Advancing Electronics Dresden cfAED, Technische Universit\"at Dresden, Germany}
\IEEEauthorblockA{\IEEEauthorrefmark{5}Institute for Integrated Signal Processing Systems, RWTH Aachen University, Germany, meyr@iss.rwth-aachen.de}
\IEEEauthorblockA{\IEEEauthorrefmark{6}GISC -- Grupo Interdisciplinar de Sistemas Complejos, Madrid, Spain}
\vspace{-0.4cm}}

\maketitle

\begin{abstract}
We provide a novel analysis of Wald's sequential probability ratio test based on information theoretic measures for symmetric thresholds, symmetric noise, and equally likely hypotheses under the assumption that the test exactly terminates at one of the thresholds. This test is optimal in the sense that it yields the minimum mean decision time. To analyze the decision-making process we consider information densities, which represent the stochastic information content of the observations yielding a stochastic termination time of the test. Based on this, we show that the conditional probability to decide for hypothesis $\mathcal{H}_1$ (or the counter-hypothesis $\mathcal{H}_0$) given that the test terminates at time instant $k$ is independent of time $k$. An analogous property has been found for a continuous-time first passage problem with two absorbing boundaries in the contexts of non-equilibrium statistical physics and communication theory. Moreover, we study the evolution of the mutual information between the binary variable to be tested and the output of the Wald test. Notably, we show that the decision time of the Wald test contains no information on which hypothesis is true beyond the decision outcome.\looseness-1 
\end{abstract}

\section{Introduction}
In many decision problems it is important to make decisions as fast as possible but with a given reliability. This problem has been first studied by A.\ Wald who introduced a
sequential probability ratio test to enable fast decisions between two possible hypotheses \cite{Wald1945}. For independent and identically distributed observations this test yields the minimum mean decision time for a decision with a given probability of error \cite{wald1948optimum}. The test accumulates the likelihood ratio given by the sequence of time discrete observations. The decision for one of the hypotheses is made as soon as the cumulative likelihood ratio reaches a given threshold which depends on the required decision reliability. The Wald test therefore corresponds to a first-passage problem with two absorbing boundaries in discrete time. The key characteristics of the Wald sequential test is that its termination time is a random quantity that depends on the actual realization of the random sequence of observations.\looseness-1

Sequential probability ratio tests can also be studied in the limit where observations occur continuously over time. In this situation, the probability ratio test becomes a continuous-time first passage problem. For continuous processes an important property of this first passage problem is that the threshold is hit exactly at a specific time. For certain systems described by a Langevin equation, it was shown previously that the probability for the process to be terminated at one of the two symmetric boundaries is independent of the time of absorption, see, e.g., \cite{LindseyMeyr77}, \cite{Roldan_etal15}.

In the present paper, we present a novel analysis of the Wald test based on information theoretic measures for symmetric thresholds, symmetric noise, and equally likely hypotheses. Under the assumption that the test exactly terminates on one of the thresholds, we show that for such a discrete-time sequential probability ratio test the probability for the test to be terminated at one of two boundaries is, as in the time continuous case, independent of the time at which the decision is taken. The assumption that the trajectory of the cumulative log-likelihood ratios exactly hits one of the thresholds is approximately fulfilled in case the average increment of the log-likelihood ratio per additional observation is small in comparison to the threshold, which holds in case the allowed error probabilities are sufficiently small. To obtain this result, we describe the behavior of the test by a recursive expression for information densities. 
This measure describes the statistical dependencies for every individual sample process. Differently, mutual information fails to analyze the realization dependent termination behavior as it takes an average over all possible realization of the observation process. Using this recursive equation we show that a key property of the Wald test is that the conditional probability to decide for hypothesis $\mathcal{H}_1$ (or the counter-hypothesis $\mathcal{H}_0$) given that the test terminates at time instant $k$ is independent of time $k$. Moreover, we show that the decision time $\tau$ of the Wald test contains no information on which hypothesis is true beyond the decision outcome. Finally, we provide an expression characterizing the evolution of the mutual information between the binary variable to be tested and the decision outcome of the Wald test.

\begin{figure*}[t!]\vspace{-0.3cm} 
	\centering \def\svgwidth{1.4\columnwidth} 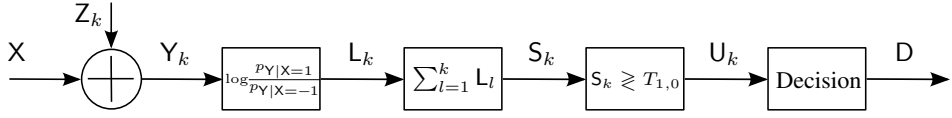\vspace{-0.1cm}
	\caption{Extended model of the Wald test}\label{Wald_ExtModel}\vspace{-0.4cm}
\end{figure*}

\section{System Model}\label{Sect_SysModel}
We consider the following decision problem on the binary random variable $\mathsf{X}\in\{-1,1\}$ based on a sequence of noisy observations of $\mathsf{X}$ where the $k$th observation is given by
\begin{IEEEeqnarray}{rCL}
\mathsf{Y}_k&=&\sqrt{\rho}\mathsf{X}+\mathsf{Z}_k, \qquad k\in\mathbb{N}.\label{Def_OutputY}
\end{IEEEeqnarray}
Here, $\mathsf{Z}_k$ is additive noise with zero mean and unit variance. The individual noise samples are assumed to be independent identically distributed (i.i.d.) with density $p_\mathsf{Z}$. In addition, we assume that the noise distribution is symmetric with respect to zero, i.e., $p_{\mathsf{Z}}(-z)=p_{\mathsf{Z}}(z)$ which is for example fulfilled by a zero-mean Gaussian distribution. Moreover, the parameter $\rho$ can be interpreted as the signal-to-noise ratio. We consider equally likely hypotheses $P(\mathsf{X}=1)=P(\mathsf{X}=-1)=\frac{1}{2}$.

The aim is to decide as fast as possible, i.e., with the lowest possible number of observations $\mathsf{Y}_k$, if $\mathsf{X}=1$ (hypothesis $\mathcal{H}_1$) or if $\mathsf{X}=-1$ (hypothesis $\mathcal{H}_0$) with a given reliability. In his seminal paper \cite{Wald1945} Wald solved this problem by providing a sequential probability ratio test (the \emph{Wald} test), which is optimal in the sense that it minimizes the mean decision time for a given reliability \cite{wald1948optimum}. The decision time $\tau$ is itself a random variable, which depends on the actual realization of the observation sequence. For this purpose, the Wald test collects observations $\mathsf{Y}_k$ until the cumulated log-likelihood ratio 
\begin{IEEEeqnarray}{rCL}
\mathsf{S}_k&=&\sum_{l=1}^{k}\mathsf{L}_k=\sum_{l=1}^{k}\log\left(\frac{p_{\mathsf{Y}|\mathsf{X}}(\mathsf{Y}_l|\mathsf{X}=1)}{p_{\mathsf{Y}|\mathsf{X}}(\mathsf{Y}_l|\mathsf{X}=-1)}\right)\label{Cumulated_LLR}
\end{IEEEeqnarray}
exceeds (falls below) a prescribed threshold $T_1$ ($T_0$). The decision time $\tau$ is the minimum value of $k$ for which $\mathsf{S}_k\notin (T_0,T_1)$.
The test decides for $\mathcal{H}_1$ ($\mathcal{H}_0$) when $\mathsf{S}_k$ first crosses $T_1$ ($T_0$), where $\mathsf{D}=1$ ($\mathsf{D}=-1$) is the decision of the test. In (\ref{Cumulated_LLR}), $p_{\mathsf{Y}|\mathsf{X}}$ denotes the probability density function of the observations $\mathsf{Y}_{k}$ conditioned on the event $\mathsf{X}$. The thresholds $T_1$ and $T_0$ depend on the maximum allowed probabilities for making a wrong decision $\alpha_1\ge P(\mathsf{D}=1|\mathsf{X}=-1)$ and $\alpha_0\ge P(\mathsf{D}=-1|\mathsf{X}=1)$. Here, $P(\mathsf{D}=1|\mathsf{X}=-1)$ ($P(\mathsf{D}=-1|\mathsf{X}=1)$) denotes the probability that the test decides for hypothesis $\mathcal{H}_1$ ($\mathcal{H}_0$) although $\mathcal{H}_0$ ($\mathcal{H}_1$) is true. The thresholds $T_1$ and $T_0$ are functions of the maximum allowed error probabilities $\alpha_1$ and $\alpha_0$. As their determination is rather involved, in the following we use the approximations $T_0\simeq\log\frac{\alpha_0}{1-\alpha_1}$ and $T_1\simeq\log\frac{1-\alpha_0}{\alpha_1}$ with $\alpha_{0},\alpha_{1}< 0.5$ \cite{Wald1945}. This choice still guarantees that the actual error probabilities are not larger than the maximum allowed error probabilities. If the test terminates exactly on one of the thresholds --- which we assume throughout this paper --- the actual error probabilities $P(\mathsf{D}=1|\mathsf{X}=-1)$ and $P(\mathsf{D}=-1|\mathsf{X}=1)$ coincide with the maximum allowed error probabilities $\alpha_1$ and $\alpha_0$. If the mean and the variance of the increments $\mathsf{L}_k$ are small in comparison to the thresholds the test ends close to one of the thresholds \cite[pp.~132-133]{Wald1945}, i.e., the assumption that the test terminates exactly on one of the thresholds is approximately fulfilled. In the present paper we focus on the important special case of $\alpha_{0}=\alpha_{1}=\alpha$, yielding symmetric thresholds $T_1=-T_0=T$, and symmetric noise.

\subsection{Three state representation of the Wald test}
\begin{figure} 
\subfigure[\hspace{-0.7cm}]
{\hspace{0.2cm}	\centering \def\svgwidth{0.35\columnwidth} \input{FigTransitionGraph_small.eps_tex}}\hspace{0.4cm}
\subfigure[\hspace{-0.3cm}]
{\hspace{-0.4cm}	\centering \def\svgwidth{0.28\columnwidth} \input{Statediagram_small.eps_tex}}
\vspace{-0.1cm}
	\caption{Trellis (a) and state transition diagram (b) of $\mathsf{U}_k$}\label{FigStatediagram}\label{FigTransitionGraph}\vspace{-0.4cm}
\end{figure}
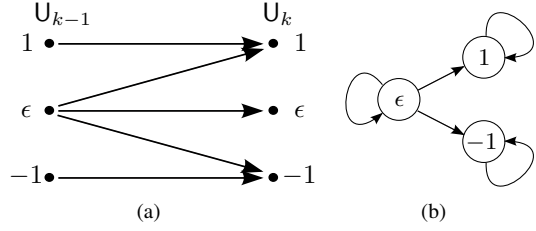

For the analysis of the Wald test we introduce the model in Fig.~\ref{Wald_ExtModel}. Here, the ternary variable $\mathsf{U}_{k}\in\mathcal{U}=\{-1,\epsilon,1\}$ with
\begin{IEEEeqnarray}{rCL}
\mathsf{U}_{k}&=&\left\{\begin{array}{ll}
1 & \textrm{if } k\ge \tau \textrm{ and }\mathsf{D}=1\\
-1 & \textrm{if } k\ge \tau \textrm{ and }\mathsf{D}=-1\\
\epsilon & \textrm{if } k<\tau
\end{array}\right. 
\end{IEEEeqnarray}
and $\mathsf{U}_{0}=\epsilon$ describes the initial state of the Wald test. The state $\epsilon$ denotes the undecided state of the test. Until $\mathsf{U}_k$ reaches $\pm1$ it is not possible to decide with the required reliability. The evolution of the state variable $\mathsf{U}_k$ can also be described by the trellis and the state transition diagram in Fig.~\ref{FigTransitionGraph}. 

To analyze the behavior of the Wald test we derive relations for the probability distribution of the state variables $\mathsf{U}_k$. In the following, we assume without loss of generality that hypothesis $\mathsf{X}=1$ is true. This is possible as we consider symmetric thresholds and a symmetric noise distribution yielding\looseness-1
\begin{IEEEeqnarray}{rCL}
P(\mathsf{U}_k=1|\mathsf{X}=1)&=&P(\mathsf{U}_k=-1|\mathsf{X}=-1)\label{SymPlusMinus}\\
P(\mathsf{U}_k=1|\mathsf{X}=-1)&=&P(\mathsf{U}_k=-1|\mathsf{X}=1)\label{SymMinusPlus}.
\end{IEEEeqnarray}

The probability that the Wald test has already made a decision for the hypothesis $\mathcal{H}_{1}$ ($\mathcal{H}_{0}$) corresponding to $\mathsf{X}=1$ ($\mathsf{X}=-1$) at the time instant $k$ or before can be expressed as
\begin{IEEEeqnarray}{rCL}
&&P(\mathsf{U}_k=a|\mathsf{X}=1)
=P(\mathsf{U}_{k-1}=a|\mathsf{X}=1)\nonumber\\
&&\qquad\quad+P(\mathsf{U}_k=a|\mathsf{U}_{k-1}=\epsilon,\mathsf{X}=1)P(\mathsf{U}_{k-1}=\epsilon)\quad\label{PUkCondX2}
\end{IEEEeqnarray}
with $a\in\{1,-1\}$. 
For (\ref{PUkCondX2}) we have used that
\begin{IEEEeqnarray}{rCL}
P(\mathsf{U}_{k-1}\!=\!\epsilon|\mathsf{X}\!=\!1)=P(\mathsf{U}_{k-1}\!=\!\epsilon)\label{PUk_cond0}
\end{IEEEeqnarray}
which follows from
\begin{IEEEeqnarray}{rCL}
&&P(\mathsf{U}_{k-1}\!=\!\epsilon)=\frac{1}{2}\!\left\{P(\mathsf{U}_{k-1}\!=\!\epsilon|\mathsf{X}\!=\!1)\!+\!P(\mathsf{U}_{k-1}\!=\!\epsilon|\mathsf{X}\!=\!-1)\right\}\nonumber\\
&&P(\mathsf{U}_{k-1}\!=\!\epsilon|\mathsf{X}\!=\!1)=P(\mathsf{U}_{k-1}\!=\!\epsilon|\mathsf{X}\!=\!-1)\label{PUk_cond0_old}
\end{IEEEeqnarray}
as we assume that both events $\mathsf{X}=1$ and $\mathsf{X}=-1$ are equally likely. Additionally (\ref{PUk_cond0_old}) follows from the assumption of a symmetric noise distribution $p_{\mathsf{Z}}$ with zero mean. Thus, (\ref{PUkCondX2}) provides a recursive relation for $P(\mathsf{U}_k=a|\mathsf{X}=1)$. The initial distribution is given by $P(\mathsf{U}_0=\epsilon)=1$ as at time instant $k=0$ no observation has been considered and the Wald test has not yet decided for one of the hypothesis. With (\ref{PUkCondX2}) we get\looseness-1
\begin{IEEEeqnarray}{rCL}
P(\mathsf{U}_k\!=\!a|\mathsf{X}\!=\!1)&=&\sum_{l=1}^{k}P(\mathsf{U}_l\!=\!a|\mathsf{U}_{l-1}\!=\!\epsilon,\mathsf{X}\!=\!1)P(\mathsf{U}_{l-1}\!=\!\epsilon)\nonumber
\end{IEEEeqnarray}
as $P(\mathsf{U}_{0}=a|\mathsf{X}=1)=0$. Note that $P(\mathsf{U}_k=a|\mathsf{X}=1)$ corresponds to the probability that the Wald test has terminated at the positive (negative) threshold \emph{at any time instant up to the time instant $k$}. On the other hand, $P(\mathsf{U}_k=a|\mathsf{U}_{k-1}=\epsilon,\mathsf{X}=1)$ is the probability that the Wald test terminates at the positive (negative) threshold \emph{at the time instant $k$}.

Finally, due to the assumption of symmetric thresholds, symmetric noise, and the assumption that $\mathsf{X}=1$ and $\mathsf{X}=-1$ are equally likely, it can be shown that for all $k$
\begin{IEEEeqnarray}{rCL}
P(\mathsf{U}_k=1)&=&P(\mathsf{U}_k=-1)\label{SymmetrieUncdonditional}\\
P(\mathsf{U}_k=1|\mathsf{U}_{k-1}=\epsilon)&=&P(\mathsf{U}_k=-1|\mathsf{U}_{k-1}=\epsilon).\label{SymmetrieUncdonditional_CondErase}
\end{IEEEeqnarray}

\section{Information Theoretic Analysis} 
We study the mutual information between the binary input $\mathsf{X}$ and the sequence of decision variables $\mathsf{U}_{0},\mathsf{U}_{1},\hdots,\mathsf{U}_{k}$ depending on the number of observations $k$ considered for decision-making so far. In this regard, we also discuss the statistical dependency of the decision time $\tau$ and $\mathsf{X}$. In the following, we denote the vector containing the sequence of decision variables up to the time slot $k$ by $\bm{\mathsf{U}}^{k}$, i.e., $\bm{\mathsf{U}}^{k}=\left[\mathsf{U}_0,\mathsf{U}_1,\hdots,\mathsf{U}_k\right]$.

Based on the chain rule for mutual information it holds that
\begin{IEEEeqnarray}{rCL}
I(\mathsf{X};\bm{\mathsf{U}}^{k})&=&\sum_{l=1}^{k}I(\mathsf{X};\mathsf{U}_{l}|\bm{\mathsf{U}}^{l-1}).\label{MutInfChainRule}
\end{IEEEeqnarray}
The summands on the RHS of (\ref{MutInfChainRule}) can be expressed by
\begin{IEEEeqnarray}{rCL}
I(\mathsf{X};\mathsf{U}_{k}|\bm{\mathsf{U}}^{k-1})&=&H(\mathsf{U}_{k}|\bm{\mathsf{U}}^{k-1})-H(\mathsf{U}_{k}|\bm{\mathsf{U}}^{k-1},\mathsf{X})\label{MuInfCond}
\end{IEEEeqnarray}
where $H(\cdot)$ denotes the (Shannon) entropy. 

As the additive noise samples $\mathsf{Z}_k$ are i.i.d.\ it holds that $p_{\mathsf{S}_k|\mathsf{S}_1,\hdots,\mathsf{S}_{k-1},\mathsf{X}}=p_{\mathsf{S}_k|\mathsf{S}_{k-1},\mathsf{X}}$. This and the fact that $\mathsf{U}_k$ does not change anymore once it reaches $\pm1$ implies that $P(\mathsf{U}_k|\mathsf{U}_1,\hdots,\mathsf{U}_{k-1},\mathsf{X})=P(\mathsf{U}_k|\mathsf{U}_{k-1},\mathsf{X})$. Likewise we get $P(\mathsf{U}_k|\mathsf{U}_1,\hdots,\mathsf{U}_{k-1})=P(\mathsf{U}_k|\mathsf{U}_{k-1})$. Hence, it holds that
\begin{IEEEeqnarray}{rCL}
H(\mathsf{U}_{k}|\bm{\mathsf{U}}^{k-1})&=&H(\mathsf{U}_{k}|\mathsf{U}_{k-1})\label{H_UcondU_1}\\
H(\mathsf{U}_{k}|\bm{\mathsf{U}}^{k-1},\mathsf{X})&=&H(\mathsf{U}_{k}|\mathsf{U}_{k-1},\mathsf{X})\label{H_UcondU_1X}
\end{IEEEeqnarray}
yielding\vspace{-0.1cm} 
\begin{IEEEeqnarray}{rCL}
I(\mathsf{X};\bm{\mathsf{U}}^{k})&=&\sum_{l=1}^{k}I(\mathsf{X};\mathsf{U}_{l}|\mathsf{U}_{l-1}).
\end{IEEEeqnarray}
Thus, the increase in mutual information between the sequence $\bm{\mathsf{U}}^{k}$ and $\mathsf{X}$ from the time $k-1$ to time $k$ is given by\looseness-1
\begin{IEEEeqnarray}{rCL}
&&I(\mathsf{X};\bm{\mathsf{U}}^k)-I(\mathsf{X};\bm{\mathsf{U}}^{k-1})=I(\mathsf{X};\mathsf{U}_k|\mathsf{U}_{k-1})\nonumber\\
&&\quad=I(\mathsf{X};\mathsf{U}_k,\mathsf{U}_{k-1})-I(\mathsf{X};\mathsf{U}_{k-1})\nonumber\\
&&\quad=I(\mathsf{X};\mathsf{U}_{k-1}|\mathsf{U}_k)+I(\mathsf{X};\mathsf{U}_{k})-I(\mathsf{X};\mathsf{U}_{k-1}).\label{VectMutInforDensChainRule}
\end{IEEEeqnarray}
where we applied twice the chain rule for mutual information.

\begin{lemma}\label{Theorem_1}
For the Wald test on the model defined in Section~\ref{Sect_SysModel}, including the assumption that the cumulative log-likelihood ratio exactly hits one of the thresholds, it holds that 
\begin{IEEEeqnarray}{rCL}
I(\mathsf{X};\mathsf{U}_{k-1}|\mathsf{U}_{k})&=&0.\label{MutInfX_Uk-1_cond_Uk_new}
\end{IEEEeqnarray}
\end{lemma}

\begin{proof}
In Appendix~A we show that
\begin{IEEEeqnarray}{rCL}
P(\mathsf{X}|\bm{\mathsf{U}}^k)&=&P(\mathsf{X}|\mathsf{U}_k).\label{OptCondCondDensity}
\end{IEEEeqnarray}
Using Bayes' rule and (\ref{OptCondCondDensity}) we obtain
\begin{IEEEeqnarray}{rCL}
P(\mathsf{X},\mathsf{U}_{k-1}|\mathsf{U}_k)&=&P(\mathsf{X}|\mathsf{U}_k)P(\mathsf{U}_{k-1}|\mathsf{U}_k).\label{Key_ProbExpr}
\end{IEEEeqnarray}
Hence, $\mathsf{X}$ and $\mathsf{U}_{k-1}$ are conditionally independent and the following Markov chain holds $\mathsf{X}\leftrightarrow\mathsf{U}_{k}\leftrightarrow\mathsf{U}_{k-1}$. Thus, (\ref{MutInfX_Uk-1_cond_Uk_new}) immediately follows.
\end{proof}

Lemma~\ref{Theorem_1} implies that $\mathsf{U}_{k}$ carries all information on $\mathsf{X}$ that is contained in $\mathsf{U}_{k-1}$. 

Using Lemma~\ref{Theorem_1} and (\ref{VectMutInforDensChainRule}) it follows that\looseness-1 
\begin{IEEEeqnarray}{rCL}
I(\mathsf{X};\mathsf{U}_k)&=&I(\mathsf{X};\mathsf{U}_{k-1})+I(\mathsf{X};\mathsf{U}_k|\mathsf{U}_{k-1}).\label{MutInforDensChainRule}
\end{IEEEeqnarray}
Eq.~(\ref{MutInforDensChainRule}) describes the increase of the mutual information between $\mathsf{X}$ and the decision variable $\mathsf{U}_{k}$ over one time step. As $\mathsf{U}_k$ can be considered as the output of the Wald test at time $k$, $I(\mathsf{X};\mathsf{U}_k)$ is the information the Wald test gives on $\mathsf{X}$ at time $k$ averaged over all possible realizations of the Wald test.

\begin{theorem}\label{Corollary_1}
Under the same conditions as in Lemma~\ref{Theorem_1} the Wald test satisfies that 
\begin{IEEEeqnarray}{rCL}
I(\mathsf{X};\tau|\mathsf{U}_k)&=&0.\label{Eq_Theorem1}
\end{IEEEeqnarray}
\end{theorem}

\begin{proof}
It holds that $P(\mathsf{X}|\bm{\mathsf{U}}^k)=P(\mathsf{X}|\mathsf{U}_k,\tau)$ as the last element of $\bm{\mathsf{U}}^k$ given by $\mathsf{U}_k$ and the decision time $\tau$ contain the same information as $\bm{\mathsf{U}}^k$, cf. Fig.~\ref{FigTransitionGraph}. Thus, with (\ref{OptCondCondDensity}) eq.\ (\ref{Eq_Theorem1}) follows immediately.
\end{proof}

Theorem~\ref{Corollary_1} states that the decision time $\tau$ gives no  additional information on $\mathsf{X}$ beyond $\mathsf{U}_k$, i.e., $I(\mathsf{X};\mathsf{U}_k,\tau)=I(\mathsf{X};\mathsf{U}_k)$.

\subsection{Information Densities}
As the mutual information by definition is an average over all involved random quantities, (\ref{MutInforDensChainRule}) does not reflect the fact that the termination time of the Wald test depends on the actual realization of the observation sequence. Differently, (\ref{MutInforDensChainRule}) describes the behavior of averages over all observation sequences $\mathsf{Y}_1,\hdots,\mathsf{Y}_{k}$. To be able to resolve the actual termination behavior, we need an expression reflecting the realization dependent termination time  corresponding to (\ref{MutInforDensChainRule}) but being more restrictive in the sense that it holds for every individual observation process. For this purpose we state the following.

\begin{corollary}\label{Corollary_2}
Under the same conditions as in Lemma~\ref{Theorem_1} the following recursive expression for information densities holds\looseness-1
\begin{IEEEeqnarray}{rCL}
i(\mathsf{X};\mathsf{U}_k)&=&i(\mathsf{X};\mathsf{U}_{k-1})+i(\mathsf{X};\mathsf{U}_k|\mathsf{U}_{k-1})\label{InforDensChainRule}
\end{IEEEeqnarray}
where the information densities are defined as $i(\mathsf{X};\mathsf{U}_k)=\log\left(\frac{P(\mathsf{U}_k|\mathsf{X})}{P(\mathsf{U}_k)}\right)$ and $i(\mathsf{X};\mathsf{U}_k|\mathsf{U}_{k-1})=\log\left(\frac{P(\mathsf{U}_k|\mathsf{U}_{k-1},\mathsf{X})}{P(\mathsf{U}_k|\mathsf{U}_{k-1})}\right)$ \cite{gray2011entropy}.
\end{corollary}
\begin{proof}
Eq.\ (\ref{InforDensChainRule}) for information densities is equivalent to
\begin{IEEEeqnarray}{rCL}
\frac{P(\mathsf{U}_k|\mathsf{X})}{P(\mathsf{U}_k)}&=&\frac{P(\mathsf{U}_{k-1}|\mathsf{X})}{P(\mathsf{U}_{k-1})}\frac{P(\mathsf{U}_k|\mathsf{U}_{k-1},\mathsf{X})}{P(\mathsf{U}_k|\mathsf{U}_{k-1})}\label{OptCondProb}
\end{IEEEeqnarray}
which follows from (\ref{Key_ProbExpr}). 
\end{proof}

Note that $I(\mathsf{X};\mathsf{U}_k)=\mathrm{E}_{\mathsf{X},\mathsf{U}_k}[i(\mathsf{X};\mathsf{U}_k)]$. Hence, we get (\ref{MutInforDensChainRule}) by taking the expectation of (\ref{InforDensChainRule}) with respect to all random quantities and, thus, (\ref{InforDensChainRule}) implies (\ref{MutInforDensChainRule}) but not vice versa. 

To describe the behavior of the Wald test to terminate when the process of cumulative log-likelihood ratios $\mathsf{S}_{k}$ reaches one of the thresholds at $\pm T$, (\ref{InforDensChainRule}) must hold for all combinations of events of $\mathsf{U}_k$, $\mathsf{U}_{k-1}$, and $\mathsf{X}$. 

\subsection{Time-Independence of Decision Probabilities}
Corollary~\ref{Corollary_2} allows to prove the following new Theorem.

\begin{theorem}\label{Theorem2}
For the system model in Section~\ref{Sect_SysModel} the following holds. In case the Wald test terminates at time $k$ the probability to decide for hypothesis $\mathcal{H}_1$ is independent of time, i.e. 
\begin{IEEEeqnarray}{rCL}
P(\mathsf{U}_{k}=1|\mathsf{U}_{k-1}=\epsilon,\mathsf{X}=1,\mathsf{U}_{k}\ne \epsilon)&=&\kappa, \quad \forall k\in \mathbb{N}\qquad
\end{IEEEeqnarray}
with the constant $\kappa=P(\mathsf{D}=1|\mathsf{X}=1)$. Equivalently, 
\begin{IEEEeqnarray}{rCL}
P(\mathsf{U}_{k}=-1|\mathsf{U}_{k-1}&=&\epsilon,\mathsf{X}=1,\mathsf{U}_{k}\ne \epsilon)=1-\kappa, \quad \forall k\in \mathbb{N}.\nonumber
\end{IEEEeqnarray} 
\end{theorem}

\begin{lemma}\label{Lemma}
The statement in Theorem~\ref{Theorem2} is equivalent to the property that the ratio of the termination probabilities at both boundaries is independent of the time instant $k$, i.e.,
\begin{IEEEeqnarray}{rCL}
\frac{P(\mathsf{U}_k=1|\mathsf{U}_{k-1}=\epsilon,\mathsf{X}=1)}{P(\mathsf{U}_k=-1|\mathsf{U}_{k-1}=\epsilon,\mathsf{X}=1)}&=&\gamma, \quad \forall k\in\mathbb{N}\label{MainKeyRatioTimeIndependent}
\end{IEEEeqnarray}
with $\gamma$ being a positive constant.
\end{lemma}

\begin{proof}[Proof of Lemma~\ref{Lemma}]
The Wald test fulfills (\ref{InforDensChainRule}) and, thus, (\ref{OptCondProb}) for all combinations of $\mathsf{X}\in\{-1,1\}$, and $\mathsf{U}_k, \mathsf{U}_{k-1}\in\mathcal{U}$. Due to the symmetry of the problem we consider only the case $\mathsf{X}=1$ w.l.o.g.. Evaluating (\ref{OptCondProb}) for different values of $\mathsf{U}_k$ and $\mathsf{U}_{k-1}$ yields\looseness-1
\begin{IEEEeqnarray}{rCL}
\frac{P(\mathsf{U}_k=1|\mathsf{X}=1)}{P(\mathsf{U}_k=1)}&=&\frac{P(\mathsf{U}_{k-1}=1|\mathsf{X}=1)}{P(\mathsf{U}_{k-1}=1)}\label{Cond1}\\
\frac{P(\mathsf{U}_k=1|\mathsf{X}=1)}{P(\mathsf{U}_k=1)}&=&\frac{P(\mathsf{U}_k=1|\mathsf{U}_{k-1}=\epsilon,\mathsf{X}=1)}{P(\mathsf{U}_k=1|\mathsf{U}_{k-1}=\epsilon)}\label{Cond2}\\
\frac{P(\mathsf{U}_k=-1|\mathsf{X}=1)}{P(\mathsf{U}_k=-1)}&=&\frac{P(\mathsf{U}_k=-1|\mathsf{U}_{k-1}=\epsilon,\mathsf{X}=1)}{P(\mathsf{U}_k=-1|\mathsf{U}_{k-1}=\epsilon)}\quad \label{Cond3}\\
\frac{P(\mathsf{U}_k=-1|\mathsf{X}=1)}{P(\mathsf{U}_k=-1)}&=&\frac{P(\mathsf{U}_{k-1}=-1|\mathsf{X}=1)}{P(\mathsf{U}_{k-1}=-1)}.\label{Cond4}
\end{IEEEeqnarray}
For (\ref{Cond1}) to (\ref{Cond4}) we have used that $P(\mathsf{U}_k\!=\!a|\mathsf{U}_{k-1}\!=\!a,\mathsf{X}\!=\!1)=1$ and $P(\mathsf{U}_k\!=\!a|\mathsf{U}_{k-1}\!=\!a)=1$ with $a\!\in\!\{1,-1\}$ and (\ref{PUk_cond0}).\looseness-1

Due to the symmetry of the test and the considered scenario reflected by (\ref{SymmetrieUncdonditional}) based on (\ref{Cond1}) and (\ref{Cond4}) we get
\begin{IEEEeqnarray}{rCL}
\frac{P(\mathsf{U}_k=1|\mathsf{X}=1)}{P(\mathsf{U}_k=-1|\mathsf{X}=1)}&=&\frac{P(\mathsf{U}_{k-1}=1|\mathsf{X}=1)}{P(\mathsf{U}_{k-1}=-1|\mathsf{X}=1)}=\gamma.\label{ConstCDF}
\end{IEEEeqnarray}
I.e., the ratio between the probability that the Wald test terminates at the positive boundary and the probability that it terminates at the negative boundary \emph{at any time instant up to the time instant $k$} is constant over $k$. We denote this constant as $\gamma$. Using (\ref{Cond2}), (\ref{Cond3}), (\ref{SymmetrieUncdonditional}), (\ref{SymmetrieUncdonditional_CondErase}), and (\ref{ConstCDF}) we get\looseness-1
\begin{IEEEeqnarray}{rCL}
\frac{P(\mathsf{U}_k\!=\!1|\mathsf{X}\!=\!1)}{P(\mathsf{U}_k\!=\!-1|\mathsf{X}\!=\!1)}&=&\frac{P(\mathsf{U}_{k}\!=\!1|\mathsf{U}_{k-1}\!=\!\epsilon,\mathsf{X}\!=\!1)}{P(\mathsf{U}_{k}\!=\!-1|\mathsf{U}_{k-1}\!=\!\epsilon,\mathsf{X}\!=\!1)}=\gamma\qquad\label{ConstPDF}
\end{IEEEeqnarray}
stating that the ratio of the termination probabilities on the positive and on the negative threshold \emph{at the time instant $k$} is also a constant independent of the time instant $k$.  
\end{proof}

\begin{proof}[Proof of Theorem~\ref{Theorem2}]
As
\begin{IEEEeqnarray}{rCL}
&&P(\mathsf{U}_{k}=1|\mathsf{U}_{k-1}=\epsilon,\mathsf{X}=1,\mathsf{U}_{k}\ne \epsilon)\nonumber\\
&&=\frac{P(\mathsf{U}_{k}=1,\mathsf{U}_{k}\ne \epsilon|\mathsf{U}_{k-1}=\epsilon,\mathsf{X}=1)}{P(\mathsf{U}_{k}\ne \epsilon|\mathsf{U}_{k-1}=\epsilon,\mathsf{X}=1)}\nonumber\\
&&=\frac{P(\mathsf{U}_{k}\!=\!1|\mathsf{U}_{k-1}=\epsilon,\mathsf{X}=1)}{P(\mathsf{U}_{k}\!=\!1|\mathsf{U}_{k-1}\!=\!\epsilon,\mathsf{X}\!=\!1)\!+\!P(\mathsf{U}_{k}\!=\!-1|\mathsf{U}_{k-1}\!=\!\epsilon,\mathsf{X}\!=\!1)}\nonumber\\
&&=\frac{1}{1+\frac{P(\mathsf{U}_{k}=-1|\mathsf{U}_{k-1}=\epsilon,\mathsf{X}=1)}{P(\mathsf{U}_{k}=1|\mathsf{U}_{k-1}=\epsilon,\mathsf{X}=1)}}=\frac{1}{1+\frac{1}{\gamma}}\label{DerivLemma-1}\\
&&=P(\mathsf{D}=1|\mathsf{X}=1)=\kappa\label{DerivLemma}
\end{IEEEeqnarray}
where for (\ref{DerivLemma-1}) we have used Lemma~\ref{Lemma}. Finally, (\ref{DerivLemma}) holds as 
\begin{IEEEeqnarray}{rCL}
&&P(\mathsf{D}=1|\mathsf{X}=1)=\sum_{k=1}^{\infty}P(\mathsf{U}_{k}=1|\mathsf{U}_{k-1}=\epsilon,\mathsf{U}_{k}\ne\epsilon,\mathsf{X}=1)\nonumber\\
&&\qquad\qquad\qquad\qquad\qquad\times P(\mathsf{U}_{k-1}=\epsilon,\mathsf{U}_{k}\ne\epsilon|\mathsf{X}=1)\nonumber\\
&&=\frac{1}{1+\frac{1}{\gamma}}\sum_{k=1}^{\infty}P(\mathsf{U}_{k-1}=\epsilon,\mathsf{U}_{k}\ne\epsilon|\mathsf{X}=1)=\frac{1}{1+\frac{1}{\gamma}}\label{DerivLemma+1}
\end{IEEEeqnarray}
where for (\ref{DerivLemma+1}) we have used (\ref{DerivLemma-1}) and the fact that the Wald test terminates almost surely \cite[Th.~6.2-1]{Melsa1978}. 
\end{proof}

Equation (\ref{MainKeyRatioTimeIndependent}) for a discrete-time problem is similar to a continuous-time result on the first passage problem with two absorbing boundaries: For stopping time distributions of stochastic entropy production an analogous expression to (\ref{MainKeyRatioTimeIndependent}) has been found for nonequilibrium steady states \cite[Eq.\ (11) and Append.~S2 in its Suppl.\ Material]{Roldan_etal15}, \cite{Neri_etal17}. Moreover, in communication theory such a symmetry has been found to show that the probability of cycle slips to the positive/negative boundary in phase-locked loops used for synchronization is independent of time \cite[Eq.\ (74)]{LindseyMeyr77}.

\subsection{Evolution of Mutual Information}
At time $k$ the mutual information between $\mathsf{X}$ and the decision variable $\mathsf{U}_k$ of the Wald test is given by
\begin{IEEEeqnarray}{rCL}
&&I(\mathsf{X};\mathsf{U}_{k})=\left\{P(\mathsf{U}_k=1|\mathsf{X}=1)+P(\mathsf{U}_k=-1|\mathsf{X}=1)\right\}\nonumber\\
&&\qquad\times\log\left(2/(P(\mathsf{U}_k=1|\mathsf{X}=1)+P(\mathsf{U}_k=-1|\mathsf{X}=1))\right)\nonumber\\
&&\qquad+P(\mathsf{U}_k=1|\mathsf{X}=1)\log\left(P(\mathsf{U}_k=1|\mathsf{X}=1)\right)\nonumber\\
&&\qquad
+P(\mathsf{U}_k=-1|\mathsf{X}=1)\log\left(P(\mathsf{U}_k=-1|\mathsf{X}=1)\right).\label{MutInf_Single2}
\end{IEEEeqnarray}
The Wald test terminates almost surely \cite[Th.~6.2-1]{Melsa1978} which means that $\lim_{k\rightarrow\infty}P(\mathsf{U}_k=\epsilon)=0$. Thus, it holds that
\begin{IEEEeqnarray}{rCL}
\lim_{k\rightarrow\infty}P(\mathsf{U}_k=-1|\mathsf{X}=1)&=&P(\mathsf{D}=-1|\mathsf{X}=1)=1\!-\!\kappa\qquad\label{LimWrong}\\
\lim_{k\rightarrow\infty}P(\mathsf{U}_k=1|\mathsf{X}=1)&=&P(\mathsf{D}=1|\mathsf{X}=1)=\kappa.\label{LimTrue}
\end{IEEEeqnarray}
Using (\ref{ConstPDF}), and $\gamma=\frac{\kappa}{1-\kappa}$, cf. (\ref{DerivLemma}), eq.\ (\ref{MutInf_Single2}) becomes
\begin{IEEEeqnarray}{rCL}
I(\mathsf{X};\mathsf{U}_{k})
&=&\frac{P(\mathsf{U}_k=1|\mathsf{X}=1)}{\kappa}I(\mathsf{X};\mathsf{U}_{\infty}).\label{EqMutInfEvol}
\end{IEEEeqnarray}
with, cf.~(\ref{LimWrong}) and (\ref{LimTrue})  
\begin{IEEEeqnarray}{rCL}
I(\mathsf{X};\mathsf{U}_{\infty})&=&1+\kappa\log(\kappa)+(1-\kappa)\log(1-\kappa).
\end{IEEEeqnarray}
I.e., $I(\mathsf{X};\mathsf{U}_{k})$ linearly increases with $P(\mathsf{U}_k=1|\mathsf{X}=1)$ until it achieves the final value $I(\mathsf{X};\mathsf{U}_{\infty})$. As we assume that the test terminates exactly on one of the thresholds it holds that $\kappa=1-\alpha$, cf.\ Sect.~\ref{Sect_SysModel}, and $I(\mathsf{X};\mathsf{U}_{\infty})$ is the mutual information to be achieved to allow a decision with the predefined error probability.\footnote{Note that (\ref{EqMutInfEvol}) does not state that $I(\mathsf{X};\mathsf{U}_{\infty})$ is only reached for $k\rightarrow\infty$.}

\section{Summary}
The analysis of the Wald test for symmetric noise, equal error probabilities corresponding to symmetric thresholds, and equally likely hypotheses provides an understanding on the implications of the information processing in optimal sequential decision-making. Under the assumption that the test terminates exactly on one of the thresholds, we mainly have shown that for these conditions (i) the decision time contains no information on which hypothesis is true beyond the decision outcome (Theorem~\ref{Corollary_1}) and that (ii) the probability to decide for hypothesis $\mathcal{H}_1$ (or $\mathcal{H}_0$) is independent of time (Theorem~\ref{Theorem2}). How far the presented results can be generalized to non-symmetric conditions will be studied in a forthcoming paper.\looseness-1

\section*{Appendix A -- Proof of (\ref{OptCondCondDensity})}
Without loss of generality let us assume that $n$ is the time instant where the state variable $\mathsf{U}_n$ changes from $\epsilon$ to $\pm 1$. If $n>k$ equation (\ref{OptCondCondDensity}) is straightforward, see Fig.~\ref{FigTransitionGraph}. For $n\le k$ and this specific realization of $\bm{\mathsf{U}}^k$ we can rewrite the LHS of (\ref{OptCondCondDensity}) as
\begin{IEEEeqnarray}{rCL} 
&&P(\mathsf{X}|\bm{\mathsf{U}}^{n-1}=\epsilon\mathbf{1}_{n-1},\bm{\mathsf{U}}_{n}^k=\pm \mathbf{1}_{k-n+1})\nonumber\\
&&\qquad=P(\mathsf{X}|\mathsf{U}_{n-1}=\epsilon,\mathsf{U}_n=\pm 1)\label{Eq_App1}
\end{IEEEeqnarray}
with $\mathbf{1}_{n}$ being the all one row vector of length $n$ and $\bm{\mathsf{U}}_{n}^k=[\mathsf{U}_n,\hdots,\mathsf{U}_k]$. Here, (\ref{Eq_App1}) follows from the fact that in case $\mathsf{U}_{n-1}$ is in state $\epsilon$ it must also have been in state $\epsilon$ in all prior time instants and once $\mathsf{U}_n$ changes to $\pm 1$ it stays in this state, cf. Fig.~\ref{FigStatediagram}. Hence, to prove (\ref{OptCondCondDensity}) it is sufficient to show that
\begin{IEEEeqnarray}{rCL} 
P(\mathsf{X}\!=\!1|\mathsf{U}_{l-1}\!=\!\epsilon,\mathsf{U}_l\!=\!1)&=&P(\mathsf{X}\!=\!1|\mathsf{U}_{l}\!=\!\epsilon,\mathsf{U}_{l+1}\!=\!1)\ \label{Eq_App2}
\end{IEEEeqnarray}
holds for an arbitrary $l$. Here, without loss of generality we assume a transition of the state variable $\mathsf{U}_l$ to $1$. Expressing (\ref{Eq_App2}) in terms of the cumulated log-likelihood ratio in (\ref{Cumulated_LLR}) yields
\begin{IEEEeqnarray}{rCL} 
P(\mathsf{X}\!=\!1|\mathsf{S}_{l-1}\!<\!T,\mathsf{S}_l\!\ge\! T)&=&P(\mathsf{X}\!=\!1|\mathsf{S}_{l}\!<\!T,\mathsf{S}_{l+1}\!\ge\! T).\hspace{0.4cm} \label{Eq_App3}
\end{IEEEeqnarray}
The LHS of (\ref{Eq_App3}) can be rewritten as follows
\begin{IEEEeqnarray}{rCL} 
&&P(\mathsf{X}=1|\mathsf{S}_{l-1}<T,\mathsf{S}_l\ge T)=\frac{P(\mathsf{X}=1,\mathsf{S}_{l-1}<T,\mathsf{S}_l\ge T)}{P(\mathsf{S}_{l-1}<T,\mathsf{S}_l\ge T)}\nonumber\\
&&=\frac{1}{1+\frac{P(\mathsf{S}_{l-1}<T,\mathsf{S}_l\ge T|\mathsf{X}=-1)}{P(\mathsf{S}_{l-1}<T,\mathsf{S}_l\ge T|\mathsf{X}=1)}}\label{Eq_App4}\\
&&=\frac{1}{1+\frac{P(\mathsf{S}_{l-1}>-T,\mathsf{S}_l\le -T|\mathsf{X}=1)}{P(\mathsf{S}_{l-1}< T,\mathsf{S}_l\ge T|\mathsf{X}=1)}}\label{Eq_App5}
\end{IEEEeqnarray}
where for (\ref{Eq_App4}) we have used that $P(\mathsf{X}=1)=P(\mathsf{X}=-1)$. Finally, (\ref{Eq_App5}) follows from the symmetry of the noise, i.e., $p_{\mathsf{Z}}(-z)=p_{\mathsf{Z}}(z)$, also cf.\ (\ref{SymMinusPlus}). Thus, to prove (\ref{Eq_App3}) we have to show that $\frac{P(\mathsf{S}_{l-1}>-T,\mathsf{S}_l\le -T|\mathsf{X}=1)}{P(\mathsf{S}_{l-1}< T,\mathsf{S}_l\ge T|\mathsf{X}=1)}$ is independent of $l$. 

The Wald test corresponds to a first passage problem of the discrete-time stochastic process $\{\mathsf{S}_1,\mathsf{S}_2,\hdots \}$ with two absorbing boundaries at $\pm T$. For a corresponding continuous-time log-likelihood ratio process $\{\tilde{\mathsf{S}}(t)\}$ we have shown the following \cite{Roldan_etal15,Neri_etal17}. Let $P(\tau;T)d\tau$ denote the probability that $\{\tilde{\mathsf{S}}(t)\}$ reaches the threshold $T$ for the first time in the time interval $[\tau,\tau+d\tau]$ given that it has not reached $-T$ before. Then it holds that \cite[Eq.\ (11) and Append.~S2 in its Suppl.\ Mat.]{Roldan_etal15}, \cite[Eq.~(E16) in Append.~E]{Neri_etal17}
\begin{IEEEeqnarray}{rCL} 
\frac{P(\tau;T)}{P(\tau;-T)}&=\exp(T)\label{ContFluctTheo}
\end{IEEEeqnarray}
i.e., the ratio between the probability that $\{\tilde{\mathsf{S}}(t)\}$ terminates in a given time interval on threshold $T$ and the probability that it terminates in the same time interval on the opposite threshold $-T$ is independent of the time $\tau$. 

Assuming that the Wald test terminates exactly on one of the thresholds, (\ref{ContFluctTheo}) implies for the discrete-time setup that (\ref{Eq_App5}) is independent of $l$ which proves (\ref{OptCondCondDensity}).

In case the trajectory of the discrete-time log-likelihood ratio process ends close to one of the thresholds, the termination behavior of the discrete-time process can be well approximated by the behavior of the continuous-time process. The assumption that the log-likelihhod process ends close to one of the thresholds is approximately fulfilled in case the size of the thresholds $T$ is large in comparison to the average increase of the log-likelihood ratio $\mathsf{L}_k$, cf.\ (\ref{Cumulated_LLR}), per observation sample. This corresponds to the case where the mean number of observations taken before making a decision is sufficiently large.\looseness-1

\section*{Acknowledgement}
This work has been partly supported by the German Research Foundation (DFG) within the Cluster of Excellence EXC 1056 'cfAED' and within the CRC 912 'HAEC'. 

\bibliographystyle{IEEEtran}
\bibliography{IEEEabrv,Bib_mod_IEEE}

\end{document}

%% file: Wald_ExtModel.eps_tex

\begingroup
  \makeatletter
  \providecommand\color[2][]{%
    \errmessage{(Inkscape) Color is used for the text in Inkscape, but the package 'color.sty' is not loaded}
    \renewcommand\color[2][]{}%
  }
  \providecommand\transparent[1]{%
    \errmessage{(Inkscape) Transparency is used (non-zero) for the text in Inkscape, but the package 'transparent.sty' is not loaded}
    \renewcommand\transparent[1]{}%
  }
  \providecommand\rotatebox[2]{#2}
  \ifx\svgwidth\undefined
    \setlength{\unitlength}{520.17786017pt}
  \else
    \setlength{\unitlength}{\svgwidth}
  \fi
  \global\let\svgwidth\undefined
  \makeatother
  \begin{picture}(1,0.1399174)%
    \put(0,0){\includegraphics[width=\unitlength]{Wald_ExtModel.pdf}}%
    \put(-0.00,0.05332806){\color[rgb]{0,0,0}\makebox(0,0)[lb]{\smash{$\mathsf{X}$}}}%
    \put(0.07042164,0.095){\color[rgb]{0,0,0}\makebox(0,0)[lb]{\smash{$\mathsf{Z}_k$}}}%
    \put(0.16155544,0.05332806){\color[rgb]{0,0,0}\makebox(0,0)[lb]{\smash{$\mathsf{Y}_k$}}}%
    \put(0.36072835,0.05332806){\color[rgb]{0,0,0}\makebox(0,0)[lb]{\smash{$\mathsf{L}_k$}}}%
    \put(0.55220474,0.05332806){\color[rgb]{0,0,0}\makebox(0,0)[lb]{\smash{$\mathsf{S}_k$}}}%
    \put(0.74368113,0.05332806){\color[rgb]{0,0,0}\makebox(0,0)[lb]{\smash{$\mathsf{U}_k$}}}%
    \put(0.94130926,0.05332806){\color[rgb]{0,0,0}\makebox(0,0)[lb]{\smash{$\mathsf{D}$}}}%
    \put(0.23053151,0.02949981){\color[rgb]{0,0,0}\makebox(0,0)[lb]{\tiny\smash{$\log\!\!\frac{p_{\mathsf{Y}|\mathsf{X}=1}}{p_{\mathsf{Y}|\mathsf{X}=\hspace{-0.1mm}-\hspace{-0.2mm}1}}$}}}%
    \put(0.42925759,0.02713843){\color[rgb]{0,0,0}\makebox(0,0)[lb]{\small\smash{$\sum_{l=1}^k\mathsf{L}_{l}$}}}%
    \put(0.61900,0.02713843){\color[rgb]{0,0,0}\makebox(0,0)[lb]{\scriptsize\smash{$\mathsf{S}_k\gtrless T_{1,0}$}}}%
    \put(0.81521034,0.02413843){\color[rgb]{0,0,0}\makebox(0,0)[lb]{\small\smash{Decision}}}%
  \end{picture}%
\endgroup

%% file: FigTransitionGraph_small.eps_tex

\begingroup
  \makeatletter
  \providecommand\color[2][]{%
    \errmessage{(Inkscape) Color is used for the text in Inkscape, but the package 'color.sty' is not loaded}
    \renewcommand\color[2][]{}%
  }
  \providecommand\transparent[1]{%
    \errmessage{(Inkscape) Transparency is used (non-zero) for the text in Inkscape, but the package 'transparent.sty' is not loaded}
    \renewcommand\transparent[1]{}%
  }
  \providecommand\rotatebox[2]{#2}
  \ifx\svgwidth\undefined
    \setlength{\unitlength}{155.45624847pt}
  \else
    \setlength{\unitlength}{\svgwidth}
  \fi
  \global\let\svgwidth\undefined
  \makeatother
  \begin{picture}(1,0.64156038)%
    \put(0,0){\includegraphics[width=\unitlength,height=2cm]{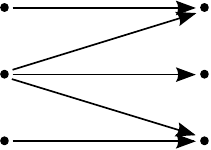}}%
    \put(-0.1,0.58){\color[rgb]{0,0,0}\makebox(0,0)[lb]{\smash{$1$}}}%
    \put(-0.1,0.3){\color[rgb]{0,0,0}\makebox(0,0)[lb]{\smash{$\epsilon$}}}%
    \put(-0.15,0.015){\color[rgb]{0,0,0}\makebox(0,0)[lb]{\smash{$-1$}}}%
    \put(-0.04623166,0.71){\color[rgb]{0,0,0}\makebox(0,0)[lb]{\smash{$\mathsf{U}_{k-1}$}}}%
    \put(0.92452056,0.71){\color[rgb]{0,0,0}\makebox(0,0)[lb]{\smash{$\mathsf{U}_k$}}}%
    \put(1.06,0.58){\color[rgb]{0,0,0}\makebox(0,0)[lb]{\smash{$1$}}}%
    \put(1.06,0.3){\color[rgb]{0,0,0}\makebox(0,0)[lb]{\smash{$\epsilon$}}}%
    \put(1.01,0.015){\color[rgb]{0,0,0}\makebox(0,0)[lb]{\smash{$-1$}}}%
  \end{picture}%
\endgroup

%% file: Statediagram_small.eps_tex

\begingroup
  \makeatletter
  \providecommand\color[2][]{%
    \errmessage{(Inkscape) Color is used for the text in Inkscape, but the package 'color.sty' is not loaded}
    \renewcommand\color[2][]{}%
  }
  \providecommand\transparent[1]{%
    \errmessage{(Inkscape) Transparency is used (non-zero) for the text in Inkscape, but the package 'transparent.sty' is not loaded}
    \renewcommand\transparent[1]{}%
  }
  \providecommand\rotatebox[2]{#2}
  \ifx\svgwidth\undefined
    \setlength{\unitlength}{142.11826148pt}
  \else
    \setlength{\unitlength}{\svgwidth}
  \fi
  \global\let\svgwidth\undefined
  \makeatother
  \begin{picture}(1,0.93232902)%
    \put(0,0){\includegraphics[width=\unitlength]{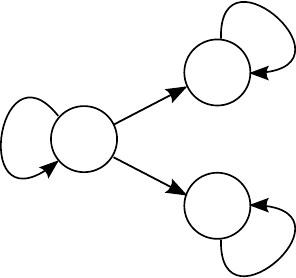}}%
    \small
    \put(0.705,0.65){\color[rgb]{0,0,0}\makebox(0,0)[lb]{\smash{$1$}}}%
    \put(0.64,0.2){\color[rgb]{0,0,0}\makebox(0,0)[lb]{\smash{$-1$}}}%
    \put(0.26,0.43){\color[rgb]{0,0,0}\makebox(0,0)[lb]{\smash{$\epsilon$}}}%
  \end{picture}%
\endgroup